\documentclass[conference]{IEEEtran}
\IEEEoverridecommandlockouts
\usepackage{cite}

\usepackage{amsmath,amssymb,amsfonts}
\usepackage{algorithmic}
\usepackage{graphicx}
\usepackage{textcomp}
\usepackage{xcolor}
\usepackage{mwe}
\usepackage{subfigure}
\usepackage{booktabs}
\usepackage{arydshln}
\usepackage{ragged2e}

\def\BibTeX{{\rm B\kern-.05em{\sc i\kern-.025em b}\kern-.08em
  T\kern-.1667em\lower.7ex\hbox{E}\kern-.125emX}}

\begin{document}
\title{MLAD: A Unified Model for Multi-system Log Anomaly Detection}

\author{
    \IEEEauthorblockN{
        Runqiang Zang $^{1}$$^{*}$\thanks{* Equal Contribution},
        Hongcheng Guo $^{2}$$^{*}$, 
        Jian Yang $^{2}$$^{+}$\thanks{+ Corresponding Authors}, 
        Jiaheng Liu $^{2}$$^{+}$, 
        Zhoujun Li $^{2}$, 
        Tieqiao Zheng$^{3}$,\\ 
        Xu Shi $^{3}$,
        Liangfan Zheng$^{3}$,
        Bo Zhang$^{3}$
    }

    \IEEEauthorblockA{$^{1}$ School of Information, Renmin University of China, Beijing, China \\ \{zangrunqiang\}@ruc.edu.cn}
    \IEEEauthorblockA{$^{2}$ State Key Lab of Software Development Environment, Beihang University \\ \{hongchengguo, jiaya, liujiaheng, lizj\}@buaa.edu.cn}
    \IEEEauthorblockA{$^{3}$ Cloudwise Research \\ \{steven.zheng, tim.shi, leven.zheng, bowen.zhang\}@cloudwise.com}

}

 
\maketitle

\begin{abstract}
In spite of the rapid advancements in unsupervised log anomaly detection techniques, the current mainstream models still necessitate specific training for individual system datasets, resulting in costly procedures and limited scalability due to dataset size, thereby leading to performance bottlenecks. Furthermore, numerous models lack cognitive reasoning capabilities, posing challenges in direct transferability to similar systems for effective anomaly detection. Additionally, akin to reconstruction networks, these models often encounter the "identical shortcut" predicament, wherein the majority of system logs are classified as normal, erroneously predicting normal classes when confronted with rare anomaly logs due to reconstruction errors.

To address the aforementioned issues, we propose MLAD, a novel anomaly detection model that incorporates semantic relational reasoning across multiple systems. Specifically, we employ Sentence-bert to capture the similarities between log sequences and convert them into highly-dimensional learnable semantic vectors. Subsequently, we revamp the formulas of the Attention layer to discern the significance of each keyword in the sequence and model the overall distribution of the multi-system dataset through appropriate vector space diffusion. Lastly, we employ a Gaussian mixture model to highlight the uncertainty of rare words pertaining to the "identical shortcut" problem, optimizing the vector space of the samples using the maximum expectation model. Experiments on three real-world datasets demonstrate the superiority of MLAD.

\end{abstract}


\section{Introduction}
\begin{figure}
\centerline{\includegraphics[width=.47\textwidth]{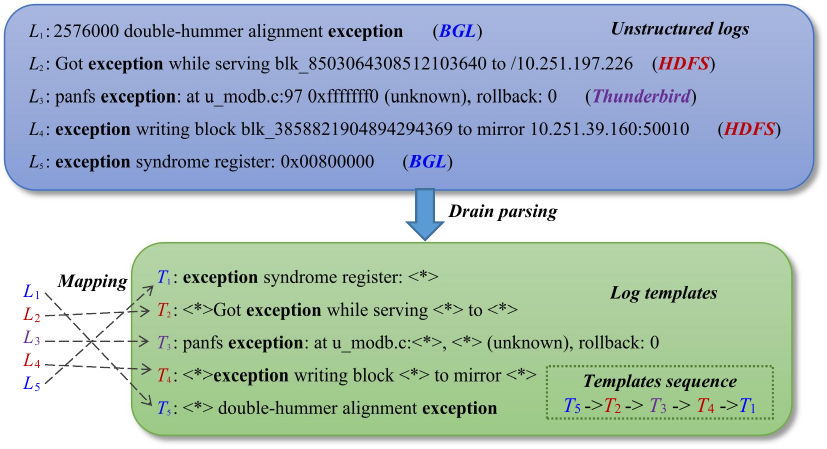}}
\caption{Log processing flow.}
\label{fig1}
\end{figure}

Log plays an essential service role in system maintenance. It records in detail the semi-structured text messages that the system has performed some operations and corresponding operation results at a specific time. In order to analyze the log, a mass amount of log messages are parsed into log templates \cite{1}, which are transformed into structured data to extract critical information. As shown in Fig.~\ref{fig1}, the log template retains the constant part of the log and ignores the unimportant variable part. The parsing still makes it easy to track possible anomalies \cite{2ZhuHLHXZL19}, such as application security, performance exceptions, diagnostic errors, and crashes.

\begin{figure*}
\centerline{\includegraphics[width=.48\textwidth]{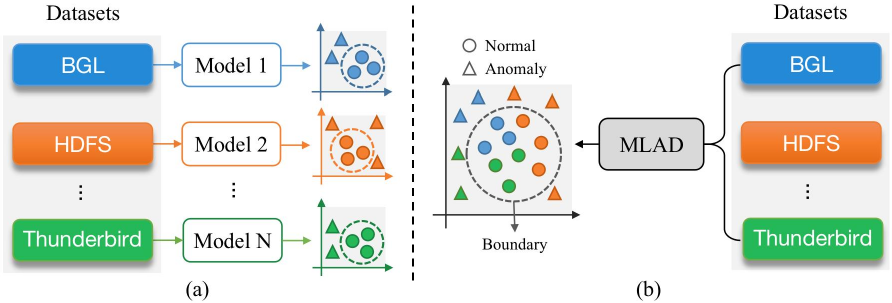}}
\caption{Multi-system log anomaly detection task. (a) Existing models learn separate decision bounds for different object logs. (b) We model the multi-system log distributions so that a single bound can detect anomalies.}
\label{fig2}
\end{figure*}

Since 2003, data-driven research efforts have been devoted to analyzing rules in log data \cite{3Du0ZS17} to detect system service anomalies automatically. Anomalies are patterns in data that do not conform to a defined notion of the system's normal behavior \cite{4ChandolaBK09}. Most anomaly detection models first train the vector space of the log template, then fit the distribution boundary of normal logs, and later detect outliers outside this distribution as anomalies\;(as shown in Fig.~\ref{fig2}). In recent years, most of the research has been conducted from a technical perspective \cite{5MengLZZPLCZTSZ19}, from the beginning based on machine learning to now based on deep learning algorithms to enhance the classification capability of the models. These studies have made very noticeable progress. However, the general approach ``one model for one system'' (Fig.~\ref{fig2}(a)) still has some limitations that have not been addressed.

\begin{enumerate}
    \item Due to the limited capacity of a single dataset, the model lacks generalization ability. By combining multiple related datasets, more data patterns and information can be captured, and anomalies that cannot be detected in a single dataset can be found.
    \item In reality, as various new systems are continuously deployed, it also takes a period of time to collect enough new logs to train the model, during which anomaly logs are easily missed. The difficulty of detecting anomalous states in new systems is exacerbated if the model lack inferential cognitive capabilities. 
    \item Few studies have taken the deep semantic features of the data as the entry point, especially in the face of log data from multiple systems, ignoring the commonality among the data from multiple systems. For example, common words in multiple systems, such as errors, exceptions, warnings, etc. and furthermore similar statement information in different system logs. 
\end{enumerate}

Therefore, to address these limitations, several perspectives are needed, such as increasing the amount of data, improving the inferential cognitive power of the model, and deeply considering the deep features of the data.
In our work, we aim to design a strongly generalized model to solve anomaly detection for multi-system logs\;(Fig.~\ref{fig2}(b)), which requires all normal samples to share the same classification boundary \cite{603687}, and the same is true for anomalous samples, so it is important to fit the distribution of normal samples in a finite vector space.
Meanwhile, the reconstruction-based model \cite{603687} is a commonly used model for anomaly detection. It trains a reconstruction model on normal samples and assumes that the reconstruction can only succeed on normal samples and will have a large reconstruction error for abnormal samples. Therefore, the reconstruction error can be used as the anomaly score. However, the reconstruction-based model encounters the problem of ``\textbf{identical shortcut}'' \cite{603687}. Although the reconstruction model is trained on normal samples, it will also reconstruct successfully when it encounters abnormal samples. It makes the reconstruction errors of both normal and abnormal samples small and difficult to be distinguished.

More importantly, based on the traditional separate setting \cite{603687,7LiSYP21} on multi-system data, the distribution of normal samples is more complicated in the unified setting as the data volume or dimension increases, exacerbating the problem of identical shortcuts. It makes the spatial vectors of normal and abnormal samples unavoidably have overlapping regions, which appear too bloated in the limited vector space \cite{4ChandolaBK09}. These overlapping regions or indistinguishable regions bring great trouble to the existing models. Therefore, we need the model to perform an appropriate deflationary transformation of the original vector space to gradually strip out the abnormal samples from the larger and more detailed vector space, focusing on allowing the model to explicitly learn the short distance relationship between similar samples and the long-distance relationship between normal/abnormal samples.

We propose a strongly generalized unified model for \textbf{M}ulti-system \textbf{L}og \textbf{A}nomaly \textbf{D}etection\;(\textbf{MLAD}) to address the above issues. The model incorporates Transformer \cite{8VaswaniSPUJGKP17} and Gaussian Mixture Model\;(GMM) \cite{9ZongSMCLCC18,10VilnisM14}, where the Transformer can efficiently learn the semantic vector space representation of multi-system logs, which consists of word-level relationship vectors generated by the trained self-attention mechanism and guidelines corrected for the induced reconstruction error, the main focus is on the keyword information in each input sequence. 
Moreover, GMM has good discrimination between normal and abnormal logs in the vector space, as reflected in the sentence-level vector space classification boundary delineation. The GMM obtains vectors from the Transformer as input. We can easily estimate the parameters of the template's GMM directly, which helps to evaluate the energy/probability of the template and output a mixed membership prediction for each word. By joint training, we minimize the encoding error from the Transformer and the template energy from the GMM, which helps the target density estimation task. The model construction idea stems from two main aspects significantly different between normal and abnormal samples:

(1)	The construction of the log vector space is different. As shown in Fig.~\ref{fig7} in Appendix, the anomaly sample vectors significantly deviate from the normal sample clusters due to the keyword information of the anomaly samples; 
(2)	The word frequency characteristics of the anomaly samples are different. Because the number of anomaly samples in the log system is usually small, and the keyword frequencies in anomaly samples are sparse. It is overall more difficult to reconstruct than normal samples. The model needs to find rare anomaly sequences that cause the system anomaly keyword information. 

To validate the effectiveness of our model, extensive experiments are conducted on the benchmark dataset and demonstrate the clear advantages of our model over previous models in three ways:
\begin{enumerate}
    \item \textbf{Anomaly detection performance:} how well the model performs in classification on each dataset.
    \item \textbf{Similar type of anomaly logs prediction:} whether the model can perform well under the fusion of multiple system datasets.
    \item \textbf{Transfer learning capability:} whether the trained model can be directly applied to the new system dataset with good capability for cross-system adaptation.
    \item \textbf{Sparse sample prediction:} whether the model can predict the logs not seen in the training dataset and the ability to mine sparse samples.
\end{enumerate}

In summary, the significant contributions of our work are as follows:
\begin{itemize}
    \item We use the bert-pooling method combination to create spatial vectors with semantic relations.
    \item We expand the vector space differentiation of self-attention by adjusting the value domain of the transformation function.
    \item Benefiting from the energy value generation of Gaussian mixture model, normal and abnormal logs have better differentiation. After applying the proposed GMM to the attention mechanism, we construct it as a weighting module of the neural network and set a specific way to maintain and normalize the basis.
\end{itemize}
The rest of the paper is organized as follows. Section~\ref{sec2} discusses the related work. Section~\ref{sec3} presents the details of MLAD. The experiment results are illustrated in Sect.~\ref{sec4}. Finally, we conclude the paper and future works in Sect.~\ref{sec5}.

\section{RELATED WORK} \label{sec2}
Traditional log anomaly detection methods mainly rely on manually setting rules or adopting statistical methods, including SVD \cite{svd},  auto-regressive integrated moving average (ARIMA) \cite{arima}, and its variants, which can meet the demand of a certain extent. However, those models are sensitive to noise and parameters \cite{wsdmDAEMON}, which limits their application in practice. In recent years, with the increasing scale of log data and the importance of log data to system operation and maintenance, log anomaly detection has become a research area of great interest \cite{cikm6,loglg,owl}. Most recent anomaly detection models have been based on deep learning networks. 

Du et al. proposed DeepLog \cite{11Du0ZS17}, a long short-term memory (LSTM) network architecture capable of identifying anomaly sequences of log service messages. For this purpose, log templates are generated, and a sequence of templates is formed as model input. The model provides a ranked output with probability for the next template in a given sequence. Then, anomaly detection is based on whether the next template has a high probability. LogAnomaly \cite{12MengLZZPLCZTSZ19} is similar to DeepLog and predicts the next log message in the sequence. Unlike DeepLog, which uses log template sequences, LogAnomaly uses log sequence embedding to improve prediction. LogTAD \cite{logtad} uses a shared LSTM model to obtain sequential representations from both the source and target systems, which is highly feasible and accurate in detecting anomalies in new systems. However, the drawback of this sequential model is that it tends to forget knowledge of important words.

The LogRobust\cite {13ZhangXLQZDXYCLC19} proposed by Zhang et al. uses an attention-based Bi-LSTM model to detect anomalies, which captures contextual information in log sequences, automatically learns the importance of different logs and can recognize and process unstable logs and sequences. LogRobust has limited ability to handle unstable logs and sequences. The LogFormer\cite{logformer} propose a pretraining and tuning pipeline to detect anomalies. Zong et al. proposed a deep autoencoding Gaussian Mixture Model\;(DAGMM) \cite{9ZongSMCLCC18} for unsupervised anomaly detection. Considering the difficulty of finding anomalous samples in the high-dimensional vector space of samples, they used a deep autoencoder to generate a low-dimensional representation, reconstruct the error for each input data point, and further input to the GMM for density estimation before performing end-to-end joint training. The situation where the model is trapped in a local optimum due to two independent steps is avoided. The disadvantage of DAGMM is the limited ability to handle large data sets.

Recent studies have shown that Transformer \cite{8VaswaniSPUJGKP17,alm,soft_template,xmt,ganlm,hanoiT,liu-etal-2022-cross-lingual,guo-etal-2022-lvp,9854132,10.1145/3583780.3614901,Wang2023RoleLLMBE,guo-etal-2023-adaptive,chen2023towards} with an attention mechanism outperforms deep learning models such as LSTM in many tasks~\cite{liu2020block}. Huang et al. \cite{14HuangLFHZYL20} used a hierarchical transformer to model log template sequences and parameter values. They redesigned a log sequence encoder and a parameter value encoder to capture log template sequences through a self-attention mechanism. Zhang et al. proposed LSADNET \cite{15ZhangWZZH21}, an unsupervised log sequence anomaly detection network based on local information extraction and global sparse Transformer models. LSADNET applies multilayer convolution to capture local correlations between neighboring logs and uses Transformer to learn global correlations between remote logs. LSADNET has limited ability to handle similar semantics between different data systems. LogBERT \cite{LogBERT} predicts masked log keys in log sequences such that normal logs are close to each other in the embedding space. CAT \cite{CAT} employs a transformer-based design of a content-aware layer that captures comprehensive event semantic information and generates semantic event representation sequences. However, these models above ignore similar semantics among multiple data systems. In contrast, our model can effectively exploit the semantic knowledge between similar logs.

\section{A Unified model for Multi-system Log Anomaly Detection} \label{sec3}
\begin{figure}
\centerline{\includegraphics[width=.45\textwidth]{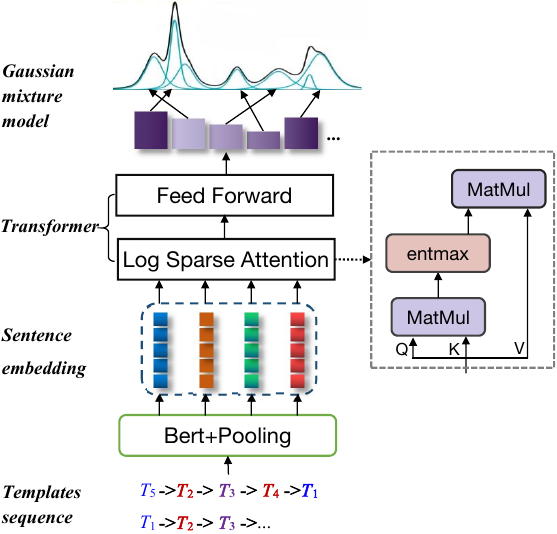}}
\caption{The architecture of our proposed MLAD.} \label{fig3}
\end{figure}

In this section, we introduce the proposed MLAD, shown in Fig.~\ref{fig3}, which combines Transformer and GMM into anomaly detection. We first describe the problem and then explain how to construct a hybrid model from log sequences and train it with unsupervised tasks to automatically find decision bounds for normal/anomalous sequence patterns.
\subsection{Problem Statement}
Given a series of unstructured system service log messages, typically containing various fields such as timestamp and severity. Thus has some sequential patterns and semantic relationships. To support downstream tasks, we follow a typical pre-processing method to represent system service log messages. We extract the log templates from log messages via the log parser Drain \cite{16HeZZL17}\;(as shown in Fig.~\ref{fig1}). For example, the BGL log template ``exception syndrome register: $<*>$'' can be extracted from the log message ``exception syndrome register: 0x008000''. $<*>$ indicates variable parameters and keeps the basic keywords. We then map each log template in the multi-system dataset to a template key: $T = \left[T_{1}, T_{2}, \ldots, T_{i}, \ldots, T_{N}\right]$, where $T_{i} \in \mathbb{T}$ denotes the template key located at the $i$th position, and $T$ denotes a set of $N$ template key sequences extracted from system service the log messages. The task objective of the model is to predict whether a new template sequence is abnormal based on a training set of log sequences containing only normal.

\subsection{Feature Extractor}
Unsupervised learning implies that the model is trained on normal log messages only. Therefore, our challenge is developing an excellent semantic understanding of normal log messages so that any significant deviation can be considered a representation anomaly. For semantic template relation learning, we use the pre-trained Sentence-bert \cite{17ReimersG19} model to obtain template sequence representations, which computes the cosine similarity between sentences to represent the spatial location relationships where their vectors are located, and after training, can effectively extract semantic representations from the original log messages, after which we Using the MEAN pooling \cite{17ReimersG19} function can losslessly compress the vectors into a template embedding of fixed dimension d to prevent information loss due to log parsing errors. These processing models facilitate the next step of single- or multi-system log fusion processing. Thus with this layer, we can obtain for each sequence $T \in \mathbb{R}^{l \times d}$ that all the template vectors are jointly formed in Large-scale high-dimensional vector spaces.

\subsection{Sparse Self-attention}

Sentence-bert only deals with sentence-level semantic relations, and its processing results are coarse. More fine-grained semantic relations require us to explore word-level spatial vectors further. We chose the self-attention mechanism, which is highly expressive and flexible for long-term and local dependencies. Self-attention encodes the input template sequence vector by associating each word in a sentence with each other based on a pairwise similarity function $f(\cdot,\cdot)$.

\begin{figure}
\centerline{\includegraphics[width=.4\textwidth]{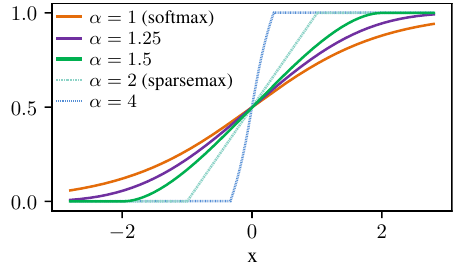}}
\caption{Illustration of entmax in the 2-dimensional
case $\alpha$-entmax.} \label{fig4}
\end{figure}

We use the linear projection $T$ to acquire the query $Q$, key $K$, and value $V$. We adopt the Scaled DotProduct Attention \cite{8VaswaniSPUJGKP17} with the sparse transformation to capture the dependency between word pairs within the sequence.
\begin{equation}
\begin{gathered}
{Q}, {K}, {V}={TW}_{{q}}, {TW}_{{k}}, {TW}_{{v}}, \\
h=\operatorname{Attention}({Q}, {K}, {V})=\alpha-\operatorname{entmax}\left(\frac{{QK}^{{\top}}}{\sqrt{{d}}}\right) {V},
\end{gathered}\label{eq1}
\end{equation}
where the learnable weights $\left\{W_{q}, W_{k}, W_{v}\right\} \in \mathbb{R}^{d \times d}, \sqrt{d_{k}}$ is a scaling factor that prevents the impact of larger values. The result $Q$ of Eq.~\ref{eq1} contains the representation matrix of the query, $K$ is the key matrix, and $V$ is the matrix of values of the terms being involved. For each sequence, the self-attention reasoning comes from the dependencies between each pair of words in the same input sequence, which means that those keywords representing normal versus abnormal log information have corresponding weights. We use a sparse transformation \cite{18PetersNM19} to increase the difference in attention weights to ensure that the embedding vector of keywords can be learned accurately. These weight values benefit from the function Eq.~\ref{eq2}, where the weight of each word is sparsely mapped to the $0$ to $1$ interval\cite{19CorreiaNM19}\;(as shown in Fig.~\ref{fig4}), and this transformation function is a core component of the self-attention mechanism:

\begin{equation}
\begin{gathered}
\alpha-\text{entmax}(x)=\underset{p \in \Delta^{d-1}}{\operatorname{argmax}} p^{\top}x+H_a^{\top}(p), \text {where} \\
H_\alpha^{\top}(p)=
\left\{\begin{array}{l} \frac{1}{\alpha(\alpha-1)}\sum_j\left(p_j-p_j^\alpha\right), \quad\alpha\neq 1 \\
H^{\top}(p), \quad\alpha=1.\end{array}\right.
\end{gathered}\label{eq2}
\end{equation}

$H_\alpha^{\top}(p)$ is Tsallis $\alpha$-entropies \cite{20Possible}, which is a family of entropies parameterized by the scalar $\alpha$ $>$ 1. From Eq.~(2) we can show that the softmax function is equivalent to 1-entmax, where Shannon entropy and Gini entropy are entropy regularizers, respectively. As shown in Fig.~\ref{fig4}, the parameter $\alpha$ controls the shape and sparsity of the function. When 1 $<$ $\alpha$ $<$ 2, the function tends to produce a sparse probability distribution with smooth corners.

If the traditional softmax function \cite{21Probabilistic} is used, the curve has a small slope at 0.5. An obvious drawback is that when the number of words is large, the weight values are dense around 0.5, which makes the differentiation of weight values between words low and finally makes it difficult for the model to identify keyword information.

\subsection{Feed-Forward Network}

Since self-attention is a linear model, we apply a fully connected Feed-Forward Network\;(FFN) to each position in the sequence to endow the model with nonlinearity and consider interactions between different latent dimensions. FFN consists of two linear transformations containing Continuously Differentiable Exponential Linear Unit\;(CeLU) \cite{27Barron17a} activation function. Given a set of vectors $[{h}_{1}, {h}_{2}, \ldots, {h}_{n}]$, the computation of a position-wise FFN sub-layer on $h$ is defined as:

\begin{equation}
\begin{gathered}
\text {FFN}(h)=\operatorname{CeLU}\left(h{W}_1+{b}_1\right) {W}_2+{b}_2, \text {where} \\
\operatorname{CeLU}(x)=\max (0, x)+\min (0, \alpha *(\exp (x / \alpha)-1),
\end{gathered}\label{eq}
\end{equation}
where $W_{1}$, $W_{2}$, $b_{1}$, and $b_{2}$ are parameters. CeLU$(\cdot)$ is the exponential linear unit activation function. We want the optimization process to be better, so we use a smoother activation function instead of a function with a less smooth transition like ReLU$(\cdot)$. It helps the model to improve its generalization ability. We also use normalization to normalize the features of each input and use dropout to avoid overfitting.
\subsection{Gaussian Mixture Model}
For the anomaly detection binary classification problem, we use the GMM to aggregate the vector space of samples, which has the advantage of optimal labeling for each sample using the pattern of the Expectation-Maximization algorithm\;(EM) \cite{22Huber2009Robust}. That is, GMM is suitable for classification problems with label-free learning. However, GMM needs to improve large-scale data handling \cite{9ZongSMCLCC18}, especially multi-system big data. In contrast, Transformer is excellent at encoding large-scale data and learning high-dimensional semantic feature representations. We can reduce the vector space dimension by adjusting the Multi-head Attention layers of the Transformer model. Therefore, combining Transformer and GMM solves the problem of GMM's weak big data processing capability.

On the other hand, Transformer's processing in binary classification still has shortcomings. After the model is trained, when the loss function is close to 0, the $\alpha$-entmax function of the attention mechanism maps the set of words in the normal log region approximately to the identity matrix. The model will have strong generalization for the word uniformity of normal logs. When the keyword weights of the abnormal logs in the test dataset are not significantly different from the normal logs, the model is prone to the ``identical shortcut'' problem, i.e., the model will misclassify the abnormal logs as normal. For this reason, we replaced the decoder component of the Transformer with GMM, which can decode the vector space more flexibly. The vector space of each sample is reconstructed iteratively by the EM method, which can increase the differentiation between normal and abnormal samples.

As the E-step in the EM algorithm, the GMM [23] prior defines the  distributions over a reconstruction function $f(h)$. Formally, the function values in any finite set of input hidden vectors should consist of integer $K$ Gaussian distributions. In general, a mixture model can use any probability distribution, and the Gaussian mixture model is used here because the Gaussian distribution has good mathematical properties and good computational performance. We first calculate the probability $\hat{\phi}_k$ that each template hidden vector $h_i$ belongs to the $k$th Gaussian distribution.
\begin{equation}
\begin{gathered}
\hat{y}=\operatorname{entmax}\left(h {W}_{h}+{b}\right), \\
\hat{\phi}_k=\sum_{i=1}^N \frac{\hat{y}_{i k}}{N},
\end{gathered}\label{eq}
\end{equation}
where $\hat{y}_{i}$ indicates the probability that the current sequence belongs to the anomaly class. It is also used as an attenuation parameter adjustment since it needs to be restricted to strictly positive values. 

Each Gaussian distribution is completely specified by its mean $\mu$ and covariance $\Sigma$. $\mu$ denotes the position in the vector space where the sample is located, and the variance $\Sigma$ is a covariance matrix. Sentence-bert calculates the similarity of templates through the cosine function, and ignores the uncertainty \cite{17ReimersG19} caused by low-frequency words when constructing word vectors, especially in the multi-system log anomaly detection scenario, the increase in the number of samples inevitably increases the imbalance of normal and abnormal samples, and the frequency of keywords in anomaly samples is relatively lower. In the above dual background, it is straightforward to cause the model to fail to learn low-frequency words adequately, triggering a relatively high uncertainty of low-frequency words and even ignoring the impact of uncommon words, eventually leading to the appearance of identical shortcut problems.

The role of the covariance matrix is to introduce the difference in uncertainty between normal and abnormal samples as part of the loss function into the anomaly detection task, increasing the ability of the model to determine the discrimination between normal and abnormal samples. The mean $\mu$ and covariance $\Sigma$ of each possible Gaussian distribution are calculated as follows.
\begin{equation}
\begin{gathered}
\hat{\mu}_k=\frac{\sum_{i=1}^N \hat{y}_{ik} h_i}{\sum_{i=1}^N \hat{y}_{i k}}, \\
\hat{\Sigma}_k=\frac{\sum_{i=1}^N \hat{y}_{ik}\left(h_i-\hat{\mu}_k\right)\left(h_i-\hat{\mu}_k\right)^\top}{\sum_{i=1}^N \hat{y}_{ik}} .
\end{gathered}\label{eq}
\end{equation}

In the following, we substitute the estimated parameters into the M-step to find the extreme value of the lower bound function, which is the derivative of the parameter and update the value of the parameter when the derivative of the lower bound function is $0$. Based on the estimated parameters, the energy of the sample can be further inferred as follows.
\begin{equation}
{E}\left(h_i\right)=-\log\left(\sum_{k=1}^K \hat{\phi}_k\frac{\exp \left(-\frac{1}{2}(h_i-\hat{\mu}_k)^{\top}\hat{\Sigma}_k^{-1}\left(h_i-\hat{\mu}_k\right)\right)}{\sqrt{\left|2\pi\hat{\Sigma}_k\right|}}\right) ,\label{eq}
\end{equation}

During the testing phase using the learned model parameters, the sample energy can be estimated directly and high-energy samples larger than the threshold are predicted as anomalies. We adjust the threshold appropriately according to the proportion of normal and abnormal samples.

\subsection{Objective Function}
Given a dataset containing N samples, the objective function guiding MLAD training is constructed as follows.
\begin{equation}
\text{Loss}=\frac{1}{{N}}\sum_{i=1}^{{N}} {L}\left({y}_i-\hat{y}_i\right)^2+\frac{\lambda_1}{{N}} \sum_{i=1}^{{N}} {E}\left(h_i\right)+\lambda_2 {P}\left(\hat{\Sigma}\right),\label{eq}
\end{equation}
where $y$ is the ground truth, $\lambda_1$ and $\lambda_2$ are the parameters that adjust the weights of each loss function. We set $\lambda_1 = 0.1$ and $\lambda_2 = -0.005$ to obtain the desired results. The whole objective function consists of three components:

${L}\left({y}_i-\hat{y}_i\right)$ is the loss function that characterizes the difference between the model predicted and true values. It is an intuitive representation of the predictive effect of the Transformer and indicates critical information about whether the model can learn the sample sequence well.
 
${E}\left(h_i\right)$ is GMM modeling the normal probability of the sample. We match the normal samples predicted by Transformer by minimizing the energy value of the normal samples, predicted by the dual decision of Transformer and GMM, and conversely, maximizing the energy value of the abnormal samples. 
 
${P}(\hat{\Sigma})$ solves the identical shortcut problem that tends to occur in log anomaly detection. Therefore, we strengthen the learning of keywords and introduce low word frequency keyword uncertainty into the loss function. The larger the uncertainty, the higher the probability of anomaly.

\section{EXPERIMENTS} \label{sec4}
In this section, we first set up the experiment. And then, we compare MLAD with the start-of-the-art baselines and analyze the results. Finally, we explore the components' role and multi-system datasets' impact on our model.
\subsection{Datasets}

We conducted experiments on BGL, HDFS, and Thunderbird datasets \cite{23OlinerS07}. The details of the three datasets are described as follows:

\textbf {BGL} is generated by the Blue Gene/L supercomputer, managed by a Machine Management Control System\;(MMCS) running on the service node, and deployed at the Lawrence Livermore National Laboratory. It contains 4,747,963 logs. Each BGL log is manually marked as normal or abnormal, and 348,460 logs are marked as abnormal.

\textbf{HDFS} consists of 11,175,629 logs collected from more than 200 Amazon EC2 nodes. Program execution in the HDFS system, e.g., writing and closing a file, usually involves a block of logs. There are 575,061 blocks of logs in the dataset, among which 16,838 blocks were labeled as anomalous by Hadoop domain experts.

\textbf{Thunderbird} is an open logs dataset collected from a Thunderbird supercomputer at Sandia National Labs. The log data contains normal and abnormal system service messages, which are manually identified. Thunderbird is a large dataset of more than 200 million log messages.

\subsection{Log Pre-processing}\label{BB}
Different datasets require pre-processing correspondingly. We extract log sequences by block IDs for HDFS since logs in HDFS with the same block ID are correlated. BGL and Thunderbird do not have such IDs, so we utilize a sliding window size is 20 without overlap to generate a log sequence. We adopt Drain \cite{16HeZZL17} with specifically designed regex to do log parsing. The number of anomalies is counted based on the window. Windows containing anomalous messages are considered anomalies. For each dataset, We choose all abnormal log sequences as the test dataset, randomly select the same number of normal log sequences as the anomalies to add to the test dataset, and the remaining normal log sequences as the training set. Table~\ref{table1} summarizes the statistics of datasets used in our experiments. The ``\# Templates'' column indicates the total number of templates in the dataset, and the number in parentheses indicates the number of templates from the test dataset that do not appear in the training dataset. The ``\# Words'' column indicates the total number of words in all templates. The ``\# Anomalies'' column indicates how many logs in the data are abnormal, which corresponds to the templates.
\begin{table}
  \caption{The Statistics of datasets}
  \label{table1}
  \centering
  \begin{tabular}{cccc}
    \toprule
    \textbf{} & BGL & HDFS & Thunderbird \\
    \midrule
    \# Log sequences & 2,780,580  &5,856,609  &9,975,120  \\
    \# Templates& 138\,(35) &44\,(25)  &1,291\,(243) \\
    \# Words & 987 &118 &6,546  \\
    \# Anomalies & 248,560  &10,109  &2,456,660 \\
    \# Train data & 2,283,460  &5,544,398 &5,061,800 \\
    \# Test data & 497,120 & 312,211  & 4,913,320 \\
    \bottomrule
  \end{tabular}
\end{table}
\section{Baselines}\label{CC}
\begin{enumerate}
    \item \textbf{DeepLog}\;(CCS-2017)\cite{11Du0ZS17}: It uses the LSTM to learn the normal execution of the system by learning the next log given the previous one. It predicts whether the incoming logs violate the predictions of the LSTM to detect anomalies.
    \item \textbf{Dagmm}\;(ICLR-2018)\cite{9ZongSMCLCC18}: A deep autoencoder Gaussian mixture model for unsupervised anomaly detection. DAGMM consists of two main components: compression and estimation networks. The compression network projects the sample into a low-dimensional space, retains the key information of the logs, and estimates the energy of the normal sample with a low score. 
    \item \textbf{LogAnomaly}\;(IJCAI-2019)\cite{12MengLZZPLCZTSZ19}: It uses log counting vectors as input to train LSTM models. They also propose template2vec, a synonym- and antonym-based model that represents log templates as semantic vectors to match new logs with existing templates.
    \item \textbf{LogRobust}\;(ESEC/FSE-2019)\cite{13ZhangXLQZDXYCLC19}: It combines a pre-trained Word2vec model with TF-IDF weights to learn a representation vector of log templates, which is then fed into an attention-based Bi-LSTM model to detect anomalies.
    \item \textbf{LogTAD}\;(CIKM-2021)\cite{logtad}: It is a cross-system log anomaly detection model based on lstm. It keeps the normal log data in the hypersphere close to the center and uses domain-adaptation to make log data from different systems follow a similar distribution.
    \item \textbf{PLELog}\;(ICSE-2021)\cite{26Yang0WWJDZ21}: It solves the under-labeling problem by probabilistic label estimation and designs an attention-based GRU neural network to detect anomalies.
    \item \textbf{LogBERT}\;(IJCNN-2021)\cite{LogBERT}:It predicts masked log keys in the log sequence so that normal log sequences are close to each other in the embedding space.
    \item \textbf{CAT}\;(KDD-2022)\cite{CAT}: It employs a self-attentive encoder-decoder transformer based architecture to learn the semantic relationships of contexts in event sequences. 
\end{enumerate}

\section{Parameters Setup}\label{DD}

We implement MLAD via Pytorch, and on three datasets, we optimize all hyperparameters of the baseline models. For a fair comparison, all models' embedding dimensions equal 100. The models are optimized using the Adam optimizer with a learning rate equal to 0.001 and a dropout rate equal to 0.5. The batch size is set to 512, and the maximum epoch is 30.

\subsection{Evaluation Metrics}\label{CC}

Anomaly detection is a binary classification problem. We label its outcome as a TP, TN, FP, and FN \cite{29icws}. TP are the anomalous logs\;(or blocks for the HDFS dataset) that are accurately determined as such by the model. TN are normal logs that are accurately determined. If the model determines a log as anomalous but normal, we label the outcome as the FP. The rest are FN. Precision, recall, and F1 score usually assess a classification model's capability \cite{wsdmmetric}.

\noindent\textbf{Precision}: Percentage of all log sequences with anomalies correctly detected by the model. $\operatorname{Precision}\,(\operatorname{Pre})=\frac{TP}{TP+FP}.$

\noindent\textbf{Recall}: Percentage of log sequences correctly identified as real overall anomalies. $\operatorname{Recall}\,
(\operatorname{Rec})=\frac{TP}{TP+FN}.$

\noindent\textbf{F1}: The summed mean of Precision and Recall. 

\noindent$\operatorname{F1}=\frac{2 \times \operatorname{Pre} \times \operatorname{Rec}}{\operatorname{Pre}+\operatorname{Rec}}.$

\begin{table*}[htbp]
\caption{The performance of different models on the three datasets, and the best model in each column is in bold.} \label{table2}
\centering
\begin{tabular}{cccccccccc}
\toprule
& \multicolumn{3}{c}{\textbf{BGL}}  & \multicolumn{3}{c}{\textbf{HDFS}}   & \multicolumn{3}{c}{\textbf{Thunderbird}}  \\
\cmidrule(lr){2-4} \cmidrule(lr){5-7} \cmidrule(lr){8-10}
            & \textbf{Pre}     & \textbf{Rec}    & \textbf{F1}  
            &  \textbf{Pre}    & \textbf{Rec}     & \textbf{F1}   
            & \textbf{Pre}    & \textbf{Rec}   & \textbf{F1}    \\ \midrule
\textbf{DeepLog}  &\textbf{0.9659}	&0.6396	&0.7696	&0.5518	&0.6785	&0.6024	&0.7538	&0.6027	&0.6699\\ 
\textbf{Dagmm}   &0.9397	&0.8831	&0.9065	&0.9018	&0.6214	&0.7358	&0.5256	&0.5395	&0.5322\\
\textbf{LogAnomaly}  &0.8918	&0.8584	&0.7428	&0.8213	&0.6179	&0.7052	&0.7672	&0.8963	&0.8273\\ 
\textbf{LogRobust}  &0.9531	&0.4766	&0.6354	&0.6989	&0.5677	&0.6700	&0.8675	&0.8652	&0.8664\\ 
\textbf{LogTAD}   &0.9102	&0.8761	&0.8949	&0.7793 &0.9091 &0.8393 &0.7523 &0.8370 &0.7886 \\
\textbf{PLELog}   &0.6843	&0.8759	&0.7314	&0.9126	&0.8373	&0.8799	&0.8606	&0.8537	&0.8671 \\
\textbf{LogBERT}   &0.8328	&0.8772	&0.8579	&0.8142	&0.7813	&0.8089	&0.8375	&0.8452	&0.8402 \\
\textbf{CAT}   &0.8727	&\textbf{0.9481}	&0.9106	&0.8638	&0.8892	&0.8771	&\textbf{0.8994}	&0.8838	&0.8923 \\

\textbf{MLAD}   &0.9492	&0.8932	&\textbf{0.9184}	&\textbf{0.9296}	&0.8656	&\textbf{0.8946}	&0.8824	&\textbf{0.9066}	&\textbf{0.8962} \\ 
\hdashline[1pt/2pt]
w/o \textbf{$\alpha$-entmax}   &0.9309	&0.8904	&0.8887	&0.7016	&\textbf{0.9773}	&0.8231	&0.7892	&0.8105	&0.8282\\ 
w/o \textbf{GMM}   &0.9128	&0.8209	&0.8644	&0.7443	&0.8131	&0.7722	&0.7534	&0.8676	&0.8053\\ \bottomrule
\end{tabular}
\end{table*}

\begin{table}
\caption{The transfer performance of the models on two similar datasets (BGL and Thunderbird).}
\label{table4}
\centering
\begin{tabular}{ccccc}
\toprule
& \multicolumn{2}{c}{\textbf{BGL→Thunderbird}}               & \multicolumn{2}{c}{\textbf{Thunderbird→BGL}}       \\ \cmidrule(lr){2-3} \cmidrule(lr){4-5} 
            & \textbf{Pre}     & \textbf{Rec}  & \textbf{Pre}     & \textbf{Rec}  \\
    \midrule
    \textbf{DeepLog}  &0.7225 &0.7368	&0.7253 &0.6817	\\ 
\textbf{Dagmm}   & 0.4998 &1.0000 & 0.5005 &1.0000	\\
\textbf{LogAnomaly}  &0.7517 &0.8602	&0.7297 &0.8029		\\ 
\textbf{LogRobust}  &0.7120 &0.8040	&0.6473 &0.9042		\\ 
\textbf{LogTAD}   & 0.8249 &0.7322	&0.7580	&0.7838   \\
\textbf{PLELog}   &0.6843	&0.7336	&0.7367	&0.7831	 \\
\textbf{LogBERT}   &0.7847	&0.7916	&0.8163	&0.8247	 \\
\textbf{CAT}   &0.7629	&0.7292	&0.8532	&0.8390	 \\
\textbf{MLAD}   &0.8277	&0.8314	&0.9404	&0.9635	 \\
    \bottomrule
  \end{tabular}
\end{table}

\subsection{Performance Comparisons}\label{DD}
In this section, we compare the overall performance of our model with some related baselines. Table~\ref{table2} shows the experimental results of all models under the Pre, Rec, and F1 evaluation indicators for three datasets. 

DeepLog focuses on considering log template sequences and ignores the semantic meaning of log templates, resulting in its poor performance on more complex datasets with many logs. In addition, DeepLog tends to mark abnormal log sequences as normal in the test dataset, generating many false predictions. logAnomaly also considers the sequential nature of logs. However, its primary benefit is using semantic vectors to match similar log templates, improving accuracy compared to DeepLog, and not marking abnormal logs as normal as DeepLog does. Moreover, LogAnomaly is more stable than other baseline models, obtaining stable F1 scores on all three datasets.
LogTAD has good performance on BGL and HDFS data, but poor performance on a larger dataset thunderbird, due to the fact that the model forgets some word information. 

Dagmm mines anomalous samples from a high-dimensional vector space by compression network, which yields good results when combined with GMM; however, compression network does not yield consistent results on all datasets, especially poor performance on the Thunderbird dataset, which has a relatively large amount of data. The main reason is that the smooth compression process loses the model from learning essential information about some keywords.

LogRobust focuses on the semantic vectors of the log templates and captures the contextual information of the semantic vectors with an attention-based Bi-LSTM. However, the main feature of the LogRobust model, which is a supervised model, requires both normal and abnormal data during the training phase, which will cost a lot of manual labeling effort. We found a performance degradation after applying LogRobust to an unsupervised task.

PLELog is a logarithm-based anomaly detection method that combines knowledge about historical anomalies through probabilistic label estimation.PLELog is immune to unstable log data through semantic embedding and detects anomalies by designing an attention-based GRU neural network. However, it is a semi-supervised model requiring some anomaly log data to train. It performs poorly on unsupervised datasets. In addition, PLELog can take a long time to train the clustering model.

Transformer-based models such as LogBERT and CAT have good anomaly detection performance, benefiting from their use of a self-attention mechanism, which can simultaneously consider all positional information in the input sequence, thus capturing more global dependencies. This modeling capability allows the model to understand the log's contextual information better. In particular, the CAT model provided a general solution to process content-rich event sequences.

In addition, the obvious drawback is that none of the baseline models can maintain a high level of performance on all three datasets. Although they can achieve very high precision or recall on some datasets, they require a balance of precision and recall performance, which results in lower F1 scores. It indicates that these models have more serious identical shortcut problems and cannot truly differentiate between normal and abnormal samples.

Our model MLAD significantly improves performance relative to all baselines. The performance advantage of MLAD is first based on the robust performance of the transformer model, which digs deep into the relationships between log template semantic vectors. From the data perspective, we also profoundly consider the significant differences between normal and abnormal samples and look for the real reasons why sample features tend to trigger model performance degradation. We found that some low-frequency keywords in the logs have to be processed. Therefore, we introduce GMM based on the transformer model to focus on the spatial location relationship of word vectors. Transformer and GMM complement each other and compensate for each other's shortcomings and solve the identical short problem. It allows the performance of MLAD to maintain a stable prediction across all three datasets, with no significant fluctuations in the results.

\subsection{Ablation Experiment}\label{EE}


We also conducted ablation experiments to examine the effect of each part of our model on the final result by removing each component. Table~\ref{table2} shows the results of the ablation experiments. The model's performance is most affected by the removal of the GMM component by MLAD, especially on the BGL and Thunderbird datasets, while there is no significant change on HDFS. Due to the number of templates in the BGL\;(138\,(35)) and Thunderbird\;(1,291\,(243)) datasets is much larger than that in HDFS\;(44\,(25)), and furthermore the number of the words that do not appear in the training dataset. It indicates a significant effect of GMM regarding the learning of sparse keywords.

\begin{figure}
  \begin{minipage}{0.46\linewidth}
    \centering
    \includegraphics[scale=0.19]{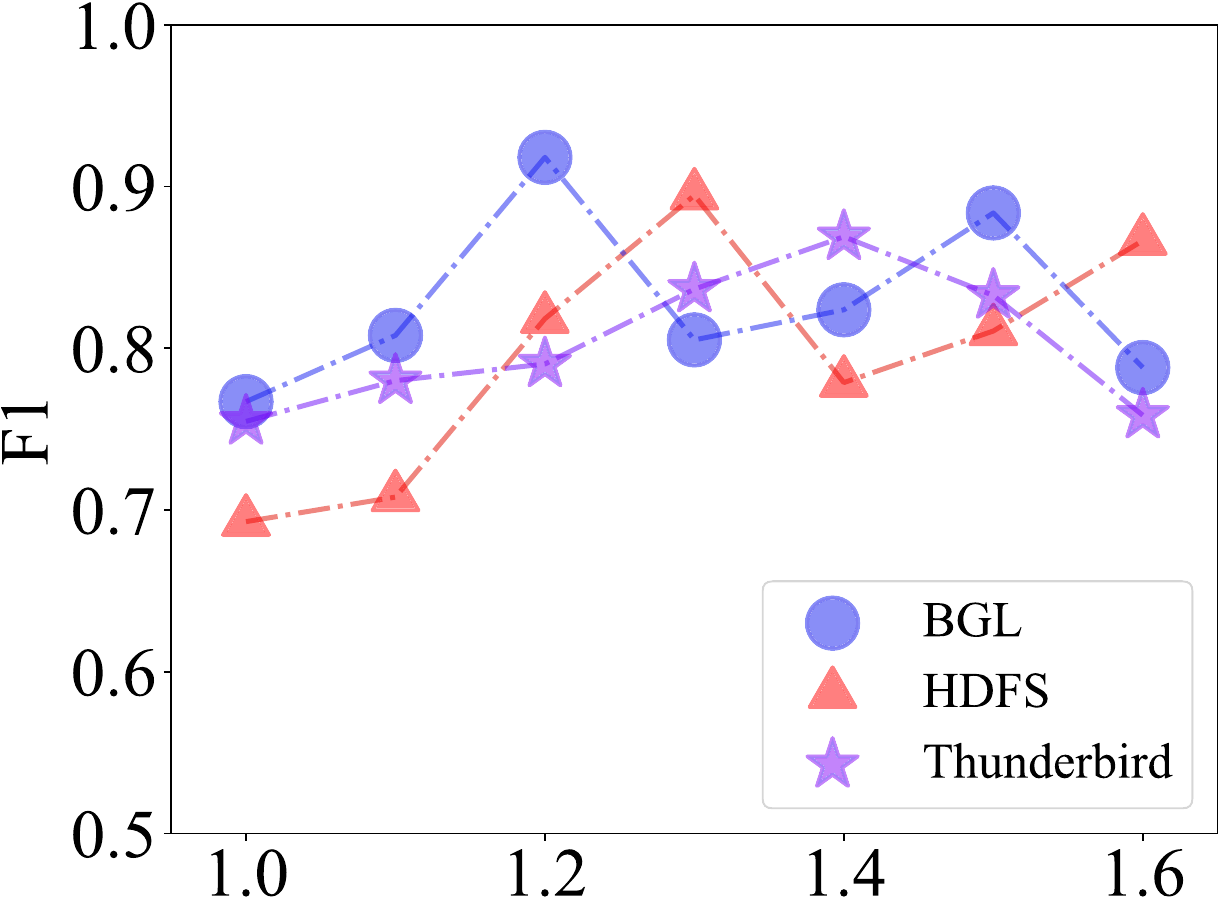}
    \caption{The effect of the $\alpha-$entmax in MLAD.}
    \label{fig5}
  \end{minipage}
  \hspace{.15in}
  \begin{minipage}{0.46\linewidth}
    \centering
    \includegraphics[scale=0.19]{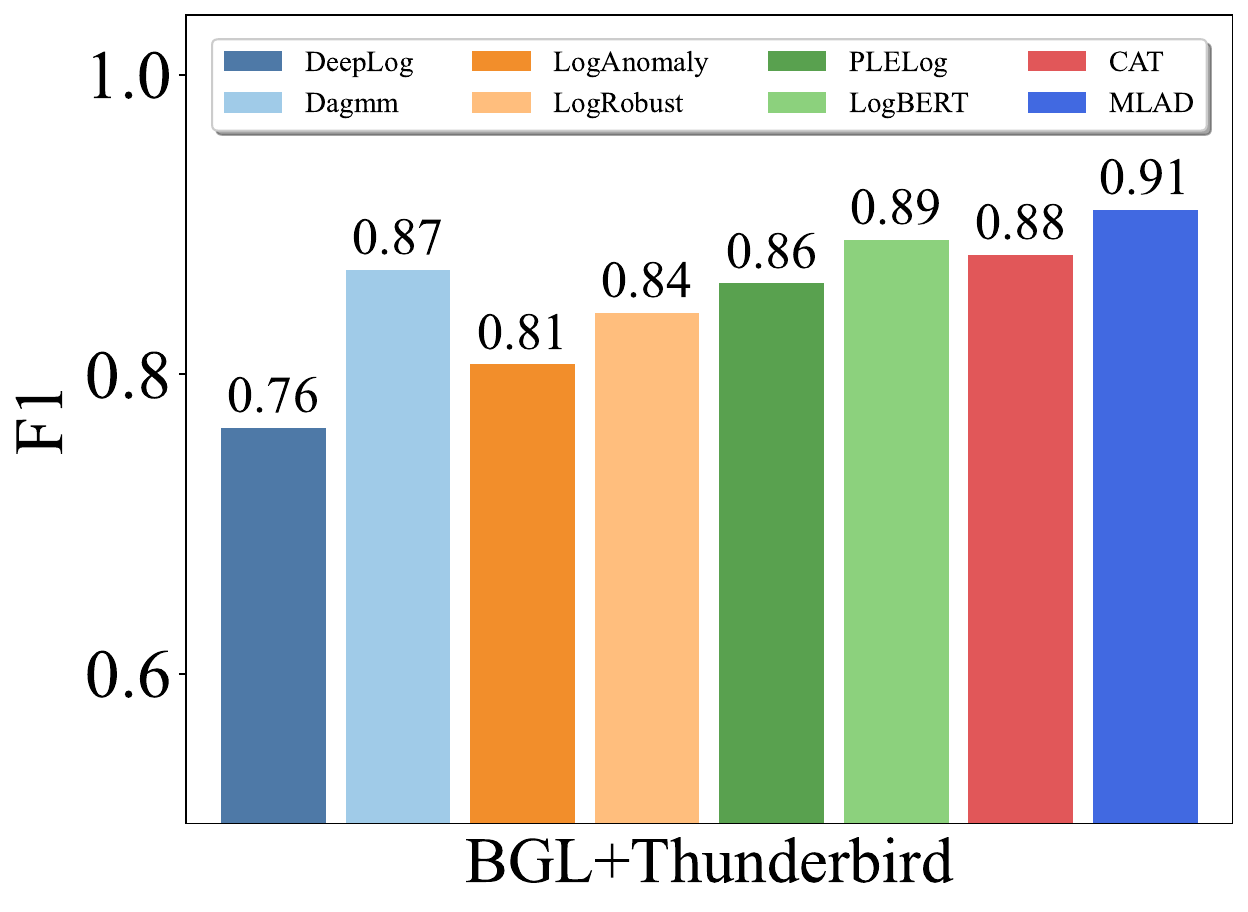}
    \caption{Experiment on multi-system datasets.}
    \label{fig6}
  \end{minipage}
\end{figure}

Removing the $\alpha$-entmax component does not significantly affect the GMM model. To demonstrate the usefulness of $\alpha$-entmax, we take the values of $\alpha$ as $\{1, 1.1, 1.2, 1.3, 1.4, 1.5, 1.6\}$ respectively. The corresponding experimental results are shown in Fig.~\ref{fig5}. In general, the performance of the softmax function is relatively general at $\alpha$ = 1. The model performs relatively well at $\alpha$ values ranging from 1.2 to 1.5. We infer that the reason is the vector space adjustment of the $\alpha$-entmax function, which sparseness the dense vector space, making it more favorable for downstream tasks and increasing the variability between normal and abnormal samples. As the value of $\alpha$ increases again, it will instead have the opposite effect, with a portion of 0-valued keyword weights generated in the sequence, causing the model to ignore some keyword weights. However, the sparse transform is generally essential to improve the prediction accuracy effectively.
\subsection{Effect on Multi-system Datasets}\label{FF}
BGL and Thunderbird datasets are pre-processed using a fixed-window mode, while HDFS datasets contain ID identifiers, and Log Pre-processing produces data of different lengths. We fuse the BGL and Thunderbird datasets as a unified dataset and experiment in the same way as a unified system. As shown in Fig.~\ref{fig6}, the experimental results show that our model detection does not degrade significantly; in contrast, the other baseline models struggle to maintain their original performance in a unified system dataset, which we conjecture is because the fused datasets require more testing of the model's generalization ability, which is something most models tend to ignore. Further, it is also possible that by merging multiple logs, the model can discover anomalies not found in individual logs. We attribute the generalization ability to MLAD's superior dual-level sentence and word feature extraction architecture. The ablation experiments show that removing the components degrades the performance to varying degrees.

First, we extract the sentence-level features of the template sequences by vectorizing the log templates with the Sentence-bert. If Sentence-bert is removed, MLAD can only learn the sequence relationships between the templates and cannot learn word-level features in the downstream task, i.e., all words in the sequence are treated equally as a bag of words. We found minimal performance degradation of the model through ablation experiments, indicating that multiple words in a template can be reduced to unique keywords. But there is a significant degradation in effect on multi-source log data. Therefore, the model ignores word-level features and only learns sentence features if it leads to insufficient generalization performance.

Second, we devise a self-attention mechanism to capture the semantic vector space representation of words in the sequence, and many similar anomalous samples consist of similar character structures. For example, although the words ``error'' and ``exception'' are different, they have similar semantics, and both indicate that the sample to which the word belongs is anomalous. In addition, the word ``exception'' followed by a series of variables such as characters or numbers should be considered to have the same semantics.

\subsection{Transfer learning between different systems}\label{FF}


In order to validate the performance of the model on cross-systems, we perform a new transfer learning experiment for the log anomaly detection task on two similar datasets: a transfer learning experiment between BGL and Thunderbird. We evaluate the performance in terms of Precision and Recall. Table~\ref{table4} shows the performance of the MLAD and baseline models. 

\textbf{(BGL→Thunderbird)} We use the BGL dataset to train the MLAD and baseline models. Then, we apply the trained models to the Thunderbird dataset for performance testing. Dagmm, DeepLog, and PLELog cannot obtain good Pre and Rec values for the unseen system dataset Thunderbird. Especially Dagmm cannot be judged as anomalous for any invisible log content, which indicates that it does not have cross-system adaptive capability. Due to effective semantic information processing, LogRobust, LogAnomaly, LogTAD, LogBERT and CAT have better transfer learning capabilities than other baseline models. Nevertheless, their performance is limited to the shared word of two similar datasets, and they need more cognitive reasoning abilities for unseen words.

\textbf{(Thunderbird→BGL)} When we set the training set as Thunderbird and the test dataset as BGL, all models have some effect improvement for the following two reasons: (1). The training set Thunderbird has more data volume compared to BGL, and the model can learn to log information more fully. (2). Although the test dataset BGL is an invisible dataset to the model, there are 261 shared words between BGL and Thunderbird, representing a larger proportion of the BGL test dataset than the Thunderbird.

Overall, better performance can be obtained for the log anomaly detection task on the new dataset using the trained MLAD, thanks to the token-by-token sequential reasoning process of the Transformer. The transfer learning experiments can demonstrate the model's robustness and generalization ability to better cope with anomaly detection tasks across different systems. 
\section{Log Vector Space Visualization}\label{GG}
\begin{figure}[htbp]
\centering    
\subfigure[LogAnomaly]{
	\begin{minipage}{0.46\linewidth}
	\centering        
	\includegraphics[scale=0.3]{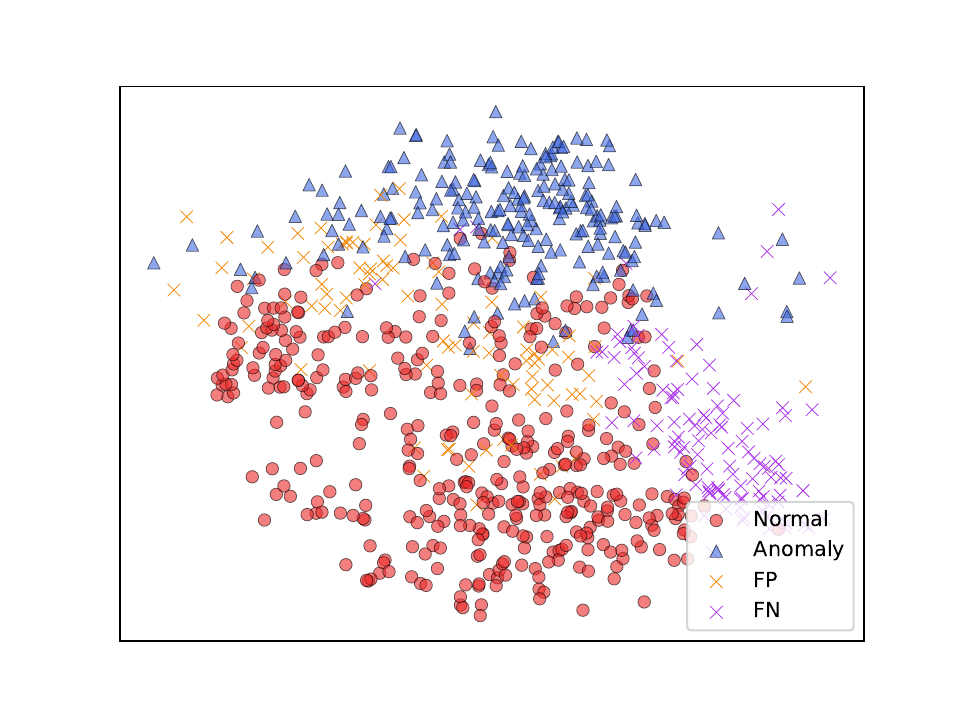}
	\end{minipage}
}
\subfigure[MLAD]{
	\begin{minipage}{0.46\linewidth}
	\centering     
	\includegraphics[scale=0.3]{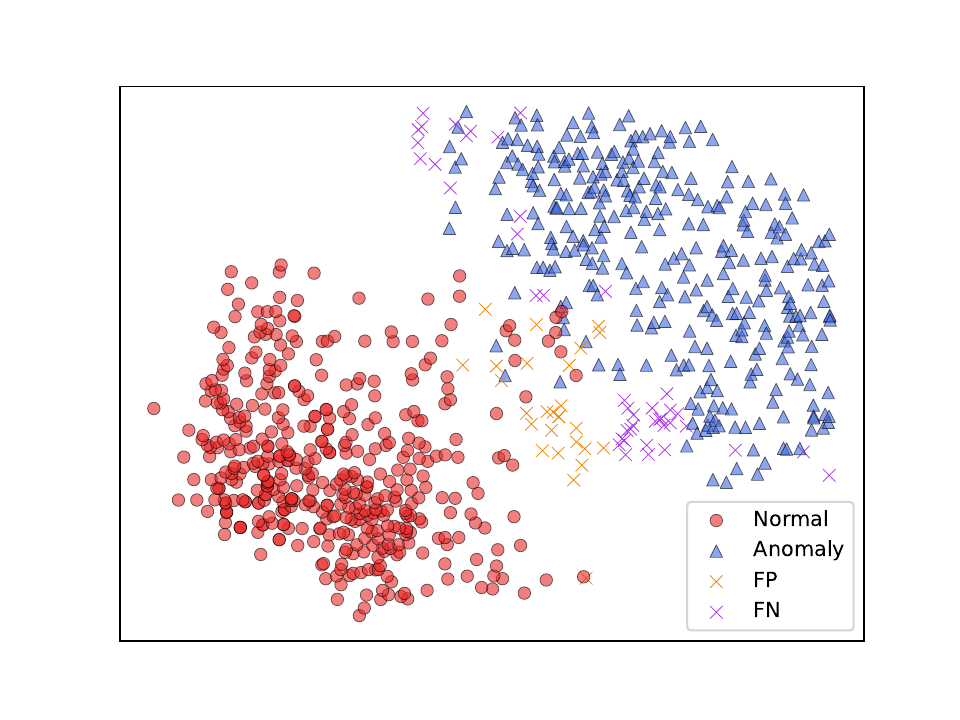}
	\end{minipage}
}
\caption{Samples in 2-dimensional space learned by LogAnomaly and MLAD. The \textcolor[RGB]{238,75,75}{red dots $\bullet$} are samples from the normal logs, and the \textcolor[RGB]{95,128,229}{blue triangles $\triangle$} are samples from the abnormal logs, the \textcolor[RGB]{242,135,5}{orange crosses $\times$}\;(FP) indicate normal samples that the model incorrectly predicts, and conversely, the \textcolor[RGB]{172,57,235}{violet crosses $\times$}\;(FN) indicate abnormal samples that the model incorrectly predicts.} \label{fig7}
\end{figure}
Because the model is trained on normal samples, it is straightforward to encounter identical shortcut problems. That is, the model easily predicts that abnormal samples are normal. We downscale the vector space of the testset to a 2-dimensional space by the t-SNE method to visually interpret the identical shortcut\;(for easy viewing, we randomly select 800 samples from the BGL testset, where the ratio of normal to abnormal samples was set to 1:1). We compare the model LogAnomaly, which also focuses on the semantic vector space. As shown in Fig.~\ref{fig7}, We can see that the visualization effect of the LogAnomaly is poor, mainly because the proportion of normal and abnormal samples determined by the model is unbalanced, or the overlapping area of normal and abnormal samples large, which makes it difficult for the model to divide the classification boundary.

In contrast, MLAD has overcome this problem. MLAD can better separate abnormal samples. A minimum number of fuzzy samples near the classification boundary are difficult to identify by the model, and the proportion of normal and abnormal samples of classification results is closest to $1:1$. It is due to the $\alpha$-entmax function in our model, which expands the distance between normal and abnormal samples in space. We can perform two-class label authentication for 2-dimensional vectors and interpret identical shortcut. Finally, combined with Table~\ref{table2}, we find that the most significant impact is the GMM component, and it is evident that the rec value drops sharply while the Pre value increases sharply. It is due to the identical shortcut phenomenon of the Transformer model after removing GMM. Under the $1:1$ ratio of normal and abnormal samples in the testset, the model defines most abnormal samples as normal. We further found that the model could not mine rare keywords, and about $30\%$ of words in the testset did not appear in the training set. In conclusion, the experiments show the superiority of our model.

\section{CONCLUSION} \label{sec5}
In this work, we propose a unified anomaly detection model MLAD for multi-system logs. Based on such a challenging task, MLAD consists of two main parts: Transformer and GMM. We have three improvements to help the model solve the common ``identical shortcut'' problem. First of all, we confirm the effectiveness of learnable semantic vector space embedding. We design a self-attention mechanism of vector space expansion to learn the internal semantic relationship of multi-system log data. Secondly, we introduce GMM to simulate the complex distribution of multi-system data. Thirdly, we focus on the uncertainty of rare keywords through the covariance of GMM, which helps the model avoid the problem of identical shortcut. Extensive experiments on three datasets and the hybrid datasets demonstrate the effectiveness of our proposed model.

In the future, we will continue to explore the area of log analysis. We hope to design a more systematic and universal log extraction method, and to obtain better
performance as much as possible on the basis of the existing
model accordingly.






\bibliographystyle{IEEEtran}
\bibliography{ref}
\end{document}